\documentclass[aps,reprint,superscriptaddress,floatfix,pra]{revtex4-2}
\usepackage{graphicx,amsmath,amssymb,amsfonts}
\usepackage{appendix,mathtools}
\usepackage{leftindex}
\usepackage[breaklinks=true,colorlinks,citecolor=blue,linkcolor=blue,urlcolor=blue]{hyperref}
\usepackage[all]{hypcap}
\usepackage{newtxtext,newtxmath}
\usepackage{natbib}
\usepackage{multirow}
\usepackage{booktabs}
\usepackage{bbold}
\usepackage{bm}
\usepackage{anyfontsize}
\usepackage{derivative}
\usepackage[dvipsnames]{xcolor}
\usepackage{ragged2e}
\usepackage{orcidlink}
\usepackage[percent]{overpic}  

\DeclareFontFamily{U}{BOONDOX-calo}{\skewchar\font=45 }
\DeclareFontShape{U}{BOONDOX-calo}{m}{n}{
  <-> s*[1.05] BOONDOX-r-calo}{}
\DeclareFontShape{U}{BOONDOX-calo}{b}{n}{
  <-> s*[1.05] BOONDOX-b-calo}{}
\DeclareMathAlphabet{\mathcalboondox}{U}{BOONDOX-calo}{m}{n}
\SetMathAlphabet{\mathcalboondox}{bold}{U}{BOONDOX-calo}{b}{n}
\DeclareMathAlphabet{\mathbcalboondox}{U}{BOONDOX-calo}{b}{n}
\begin{document}
\title{Noise-induced Zeno-like effect in a spin-chain quantum battery}
\author{Maulana M. Fajar\orcidlink{0009-0006-8496-3470}}\email[Corresponding author: ]{fajar.060503@gmail.com}
\affiliation{Research Center for Quantum Physics, National Research and Innovation Agency (BRIN), South Tangerang 15314, Indonesia}
\affiliation{Department of Physics, Sultan Maulana Hasanuddin Banten State Islamic University, Serang 4217, Indonesia}
\author{Beta N. Pratiwi\orcidlink{0000-0002-7334-7756}}
\affiliation{Department of Physics, Sultan Maulana Hasanuddin Banten State Islamic University, Serang 4217, Indonesia}
\author{Subur Pramono\orcidlink{0000-0001-6346-955X}}
\affiliation{Department of Physics, Sultan Maulana Hasanuddin Banten State Islamic University, Serang 4217, Indonesia}
\author{Gagus~K.~Sunnardianto\orcidlink{0000-0002-2781-0076}}
\affiliation{Research Center for Quantum Physics, National Research and Innovation Agency (BRIN), South Tangerang 15314, Indonesia}
\author{Ahmad~R.~T.~Nugraha\orcidlink{0000-0002-5108-1467}}\email[Corresponding author: ]{ahma080@brin.go.id}
\affiliation{Research Center for Quantum Physics, National Research and Innovation Agency (BRIN), South Tangerang 15314, Indonesia}
\affiliation{Engineering Physics Study Program, School of Electrical Engineering, Telkom University, Bandung 40257, Indonesia}
\begin{abstract}
Quantum batteries, which are energy-storage or state-storage devices that exploit unique quantum effects, are sensitive to environmental noise.  Here, we demonstrate that suitably engineered noise can induce a ``Zeno-like" stabilization effect of the charging process in a spin-chain quantum battery within the Heisenberg $XYZ$ model.  Focusing on a system size $N=6$, which balances computational cost and storage capacity, we find that the ergotropy-to-energy ratio $\mathcal{W}(t)/E_B(t)$ attains a maximum value of about $0.99$ at a certain time parameter value.  Then, by varying the noise strength in each channel, we find that not only does decoherence merely degrade performance but it may also stabilize stored energy and ergotropy in the high noise strength regime.  For instance, the phase-flip channel slows charging and reduces charging power, but its discharging behavior releases energy and ergotropy more slowly, allowing the battery to be used for longer times compared to bit-flip and bit-phase-flip channels. In contrast, the bit-flip channel enables fast charging, but yields low storage and rapid energy release.  Remarkably, the bit-phase-flip channel can combine the advantages of both bit-flip and phase-flip channels in the high-noise-strength regime.  The bit-phase-flip channel supports accelerated charging together with enhanced storage capacity, while its discharging behavior resembles that of the bit-flip channel with rapid energy release.  These results reveal that, under sufficiently strong noise, environmental decoherence induces a Zeno-like stabilization, allowing it to achieve \emph{enhanced} charging performance and to \emph{stabilize} stored energy and ergotropy in the spin-chain quantum battery.
\end{abstract}
\date{\today}
\maketitle
\section{Introduction}
\label{sec:int}
Technological revolutions often historically followed new scientific theories.  The first industrial revolution, for example, was the result of advances in thermodynamics research~\citep{carnot1824reflections, fermi1956thermodynamics}.  Classical thermodynamics sets strict limits: heat cannot flow from cold to hot, and no heat engine can exceed Carnot efficiency.  An early work on quantum thermodynamics even hypothesized that quantum entanglement alone cannot extract additional energy from a reservoir~\citep{hovhannisyan2013entanglement}, suggesting that there is no quantum advantage in equilibrium thermodynamics.  Later theoretical developments somehow overturned this view.  Creating entanglement in many‑body quantum systems enables faster work extraction~\citep{hovhannisyan2013entanglement, Gyhm2024}, sparking interest in quantum heat engines~\citep{kieu2004second, uzdin2016equivalence}, and particularly in quantum batteries, whose performance can be boosted by unique quantum effects~\citep{alicki2013entanglement,binder2015quantacell, campaioli2017quantum}.  The growing demand for energy-efficient storage is now driving research into these devices because classical batteries face inherent limits, whereas quantum batteries promise faster and denser energy storage and transfer by exploiting quantum mechanics, not only for conventional charge and energy transfer but also for a general quantum state transfer~\citep{hadipour_enhancing_2024}.  

Given these theoretical advancements, the following question naturally arises: can inherent quantum properties, such as coherence~\citep{baumgratz2014quantifying} and entanglement~\citep{horodecki2009quantum}, also enhance the efficiency of energy storage and extraction?  In this context, two main approaches to quantum batteries have been proposed.  The first involves a collection of independent quantum systems acting as battery cells, where energy is deposited and extracted through global unitary or non-unitary operations that may generate entanglement during the charging process~\citep{alicki2013entanglement, hovhannisyan2013entanglement, binder2015quantacell, campaioli2017quantum}.  This model has been widely explored and has demonstrated how collective quantum effects can speed up energy transfer.  The second approach considers using the ground state of an interacting (e.g., spin) system as the initial state of the battery.  In this model, energy is stored by applying operations that are allowed within quantum mechanics~\citep{andolina2018charger, le-spinchain-2018}.  Recent studies have shown that the nature of the interactions in the initial Hamiltonian plays a crucial role in determining the charging performance and power output of the battery~\citep{campaioli2017quantum}.  In addition to quantum coherence, charging protocols based on quantum superpositions of trajectories have been introduced~\citep{lai2024quick, Murphy2025Ergotopy}.  In some examples of this scheme, a qubit battery can interact simultaneously with multiple cavity chargers or a single charger at different entry positions.  Coherence-driven dynamics was shown to enhance the maximum extractable work (or ``ergotropy") and, in certain configurations, induce Dicke-type interference that allows for perfect charging, where all stored energy can be fully converted into work~\cite{lai2024quick, Beder2025Work}.

Recent progress in quantum technologies has also generated considerable interest in the realization of quantum batteries and thus broadened their design space~\citep{binder-qbattery-2015, campaioli2017quantum}. Among the various architectures explored, spin-chain quantum batteries are attractive due to their ease of processing, scalability, and strong interspin interactions~\citep{le-spinchain-2018, ferraro-collectivecharging-2018, Song2024Remote, Shukla2025Optimizing}.  However, real quantum systems are inherently open and interact with their environment, leading to decoherence and noise~\citep{breuer2007theory}.  In certain regimes, such interactions can suppress coherent energy transfer via mechanisms similar to the quantum Zeno effect, leading to Zeno-like behavior that reshapes charging performance~\citep{ghosh2021fast, Song2025SelfDischarging}.  In its original meaning, the quantum Zeno effect refers to the slowing or even freezing of quantum evolution under frequent measurements, effectively keeping the system in its initial state~\citep{misra1977zeno, breuer2007theory}.  In open systems, similar phenomena can arise due to strong system-environment interactions, often referred to as Zeno-like behavior or environment-induced Zeno effect~\citep{rivas2012open}.  This noise, typically regarded detrimental, has recently been considered to play a constructive role when properly engineered.  For example, a recent study proposed a universal method for fast charging using a controlled pure dephasing of the charger~\citep{Shastri2025}.  Although weak dephasing allows underdamped oscillations of stored energy and strong dephasing induces Zeno freezing, an optimal intermediate dephasing rate maximizes charging speed and robustness against detuning.  Another study demonstrated a modulator-assisted local control scheme for quantum batteries in a charger-mediated protocol~\citep{Xie2025}. By applying repeated unitary pulses to a third-party modulator system, the study leveraged the quantum Zeno effect to dynamically control the effective coupling strength between the charger and battery, enabling remote and platform-independent charging speed regulation.  The problem now is that none has considered the Zeno-like effect in spin-chain quantum battery systems.

\begin{figure*}[!t]
    \centering
        \includegraphics[width=0.97\textwidth, trim=0 40 0 40, clip]{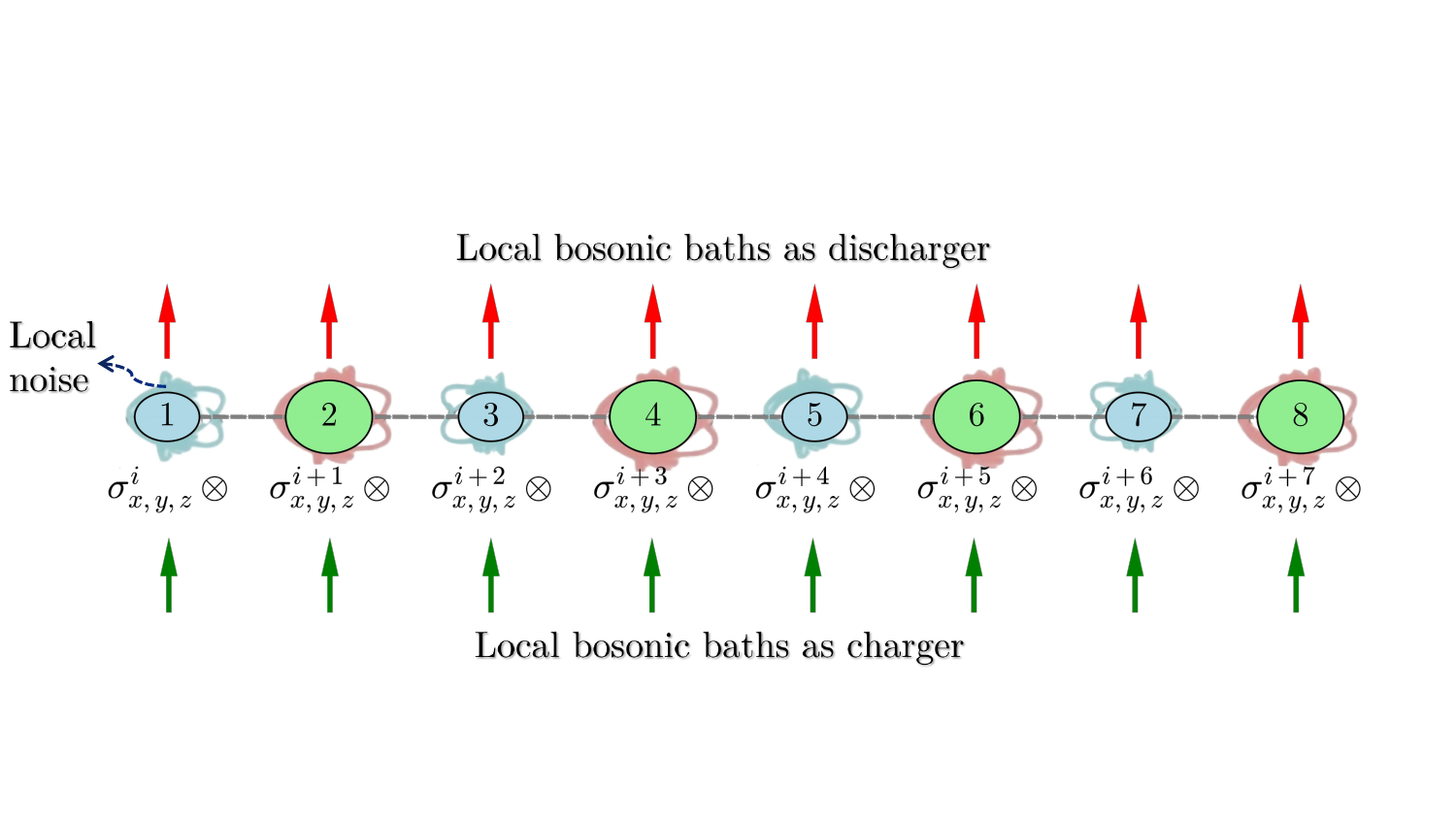}
        \caption{\justifying   Schematic illustration of the quantum battery model based on an open spin-1/2 chain with nearest-neighbor interactions. Each spin at site $i$ is coupled to local bosonic baths that act as chargers (green arrows, via the collapse operator $L_+ = \sqrt{\Gamma_+}\sigma_+^i$) or as energy extractors during harging (red arrows, via $L_- = \sqrt{\Gamma_-}\sigma_-^i$). The rates are scaled by the characteristic energy scale $\Delta E = E_{\max} - E_{\min}$, defined as the energy difference between the highest and lowest eigenvalues of the battery Hamiltonian. The charging process is therefore driven by a combination of the local bosonic baths and the charging Hamiltonian $\hat{H}_c$ defined in Eq.~\eqref{eq:c}. All parameters in the model, including $\hat{H}_c$ and the noise-channel strengths, are expressed in dimensionless form by normalizing with respect to this energy scale. In addition, each spin is subject to local noise channels corresponding to bit-flip ($\sigma_x^i$), phase-flip ($\sigma_z^i$), and bit-phase-flip ($\sigma_y^i$) operations.
    }
        \label{fig:bat}
\end{figure*}

In this work, we investigate Zeno-like effects in quantum batteries by considering a Heisenberg $XYZ$ spin chain under a transverse magnetic field with open boundary conditions~\citep{Ghosh2022, Verma2025}. Each spin is coupled to two independent local bosonic reservoirs: one responsible for charging and the other for discharging (see Fig.~\ref{fig:bat})~\citep{ghosh2021fast}.  The charging process is driven by a time-dependent Hamiltonian in combination with a local thermal bath, forming a cyclic charge–discharge protocol. To study the influence of environmental effects on energy transfer, we applied local decoherence noise to selected spins.  Within the framework of open Markovian quantum systems, battery performance is evaluated in terms of stored energy and ergotropy.  The impact of local decoherence is systematically analyzed by varying the noise strength.  We consider three canonical quantum error models: (1) bit-flip channels, (2) phase-flip channels, and (3) bit-phase-flip channels~\citep{nielsen-chuang-2010}.  These noise channels are implemented locally and act independently in each cycle~\citep{breuer2007theory, rivas2012open}.

Among the three noise channels, bit-flip and phase-flip cases generally degrade the performance of the quantum battery: even when energy saturation is reached, the accumulated energy remains lower than in the ideal, noiseless case.  In contrast, the bit-phase-flip channel exhibits qualitatively different behavior.  At low decoherence strengths, it initially hinders energy storage, but as the noise strength increases beyond a certain threshold, it accelerates charging and stabilizes both the stored energy and the extractable work. This counterintuitive effect, reminiscent of a Zeno-like suppression mechanism, demonstrates that moderate decoherence can paradoxically \emph{enhance} the charging process while simultaneously \emph{stabilizing} the energetic output of the battery over long periods.  Moreover, the impact of noise is closely intertwined with the internal coupling strength $J$ in the Heisenberg $XYZ$ spin-chain model, which governs energy transport and collective dynamics~\citep{campaioli2017quantum,andolina2018charger}.  Certain regimes of $J$ may not only mitigate the detrimental effects of decoherence but also further improve charging efficiency, suggesting a rich interplay between interaction strength and environmental noise. Although a systematic exploration of these regimes is beyond the scope of the present study, we highlight them as a promising direction for future research aimed at optimizing quantum battery performance.

The rest of this paper is organized as follows. I n Sec.~\ref{sec:model-methods}, we introduce our main model: the $XYZ$ Heisenberg spin-chain quantum battery, formulated within an open quantum system framework including the relevant noise channels.  Using the QuTiP package~\citep{johansson-qutip-2012} along with our custom code, we compute key quantum battery metrics such as stored energy, ergotropy, charging power, and charging time.  Section~\ref{sec:result} presents the simulation results, including the scaling behavior under local noise, the time evolution of energy and ergotropy during charging and discharging, and the dependence on interaction strength.  Finally, Sec.~\ref{sec:conclusion} summarizes our main findings and discusses potential implications for quantum battery design.

\section{Model and methods}
\label{sec:model-methods}
We study how environmental effects influence energy storage and extraction in a quantum battery modeled as an interacting Heisenberg $XYZ$ spin chain under a magnetic field, described by the Hamiltonian $\hat{H}_0$~\citep{Heisenberg1928, Ghosh2022}.  The system consists of $N$ spin-1/2 particles with open boundary conditions,  where each spin occupies a distinct site $i$ and interacts with its nearest neighbors. The ground state or the canonical thermal equilibrium state is considered as a potential initial state for the battery.  With such considerations, $\hat{H}_0$ can be written as
\begin{widetext}
\begin{align}
    \hat{H}_0 = \underbrace{\frac{\mathcalboondox{h}}{2} \sum_{i=1}^{N} \sigma^{i}_z}_{H_{\text{field}}} + \underbrace{\frac{J}{4} \sum_{i=1}^{N-1} \left[ (1+\gamma)\sigma^i_x \otimes \sigma^{i+1}_x 
    + (1-\gamma) \sigma^i_y \otimes \sigma^{i+1}_y \right] + \frac{J_z}{4} \sum_{i=1}^{N-1} \sigma^i_z \otimes \sigma^{i+1}_z}_{H_{\text{int}}},
    \label{eq:HB}
\end{align}
\end{widetext}
where~$\sigma_\alpha~(\text{with }\alpha=x,y,z)$ denotes the standard Pauli spin matrices, $\mathcalboondox{h}$ represents the energy of the external magnetic field applied uniformly at each site and $\gamma = [0,1]$ characterizes the degree of anisotropy in spin interactions. The parameters \( \{J\}, \text{ and } \{J_{z}\} \) in Eq.~\eqref{eq:HB} denote the nearest-neighbor coupling strengths in the $xy$-plane and along the $z$-axis, respectively, and may vary from site to site.  In the thermodynamic limit, the model exhibits a quantum phase transition at $\lambda \equiv J/\mathcalboondox{h} = 1$.  The system is in a paramagnetic phase for $|\lambda| \leq 1$, antiferromagnetic for $\lambda > 1$, and ferromagnetic for $\lambda < -1$~\citep{Sachdev2011}.

We could trivially increase the battery efficiency simply by multiplying $\hat{H}_0$ by a constant greater than one or increasing the strength of the local field component ($H_{\text{field}}$) of the Hamiltonian. To avoid such artificial enhancements and ensure a meaningful analysis, we normalize $\hat{H}_0$ as~\citep{Ghosh2020}
\begin{align}
\frac{(\hat{H}_0-E_{\min}\mathbb{1})}{(E_{\max} - E_{\min})} \longrightarrow \hat{H}_0.
\end{align}
Here, $\Delta E = E_{\max} - E_{\min}$, where $E_{\min}$ and $E_{\max}$ are the minimum and maximum eigenvalues of $\hat{H}_0$, respectively. After this transformation, the spectrum of $\hat{H}_0$ is always confined within the interval $[0,1]$.  The parameter $\Delta E$ thus provides a natural energy scale for the system, and in the following analysis all quantities with dimensions of energy will be expressed in units of $\Delta E$.  This bounded spectrum allows us to clearly assess the contribution of the interaction term $H_{\text{int}}$ to the charging power, in contrast to the baseline case where $H_{\text{int}} = 0$, which may not exhibit any quantum features~\citep{Ghosh2020}.

During the charging stage, the quantum battery, modeled as an interacting spin system, is coupled to a composite charger consisting of two components: (i)~a set of local bosonic reservoirs and (ii)~a coherent external drive represented by the charging Hamiltonian. The bosonic reservoirs act as absorption channels that mediate the transfer of energy into the system, and after tracing out their degrees of freedom, their effect can be described using an effective dynamical map. All coupling strengths and energy parameters are expressed in units of the normalized energy scale $\Delta E$. At the same time, the coherent energy input is governed by
\begin{align}
\hat{H}_c = \frac{\omega}{2} \sum_{i=1}^N \sigma_x^{i}, \label{eq:c}
\end{align}
where $\omega$ denotes the strength of the applied local external magnetic field in the $x$-direction. The above Hamiltonian is expressed in units of the normalized energy scale $\Delta E$, so that $\omega$ is understood as a dimensionless parameter. This charging Hamiltonian describes the effect of a transverse magnetic field applied locally to each spin, which allows coherent excitation during the charging process. The combined action of the Hamiltonian and absorption channels drives the quantum battery from its initial state to a higher energy configuration. Charging is considered complete once the stored energy stabilizes and no longer changes over time.

During this evolution, the system may also be subject to environmental noise such as bit-flip, phase-flip, and bit-phase-flip errors, which can affect the efficiency and stability of energy storage. The open-system dynamics of the quantum battery, including both the unitary evolution generated by $\hat{H}_c$ and the dissipative effects from reservoirs and noise, are modeled using the Gorini-Kossakowski-Sudarshan-Lindblad master equation formalism~\citep{breuer2007theory, gorini1976completely, lindblad1976generators}.
\begin{align}
\partial_t \rho(t) &= -i [\hat{H}(t), \rho(t)] \notag \\
& \quad + \sum_{k} \Gamma_{(k)} \left( \hat{L}_{k,i} \rho(t) \hat{L}_{k,i}^{\dagger} - \frac{1}{2} { \hat{L}_{k,i}^{\dagger} \hat{L}_{k,i}, \rho(t) } \right).
\label{eq:lindblad}
\end{align}
The Lindblad rates $\Gamma_{(k)}$ are expressed in units of inverse time so that the associated Lindblad operators $\hat{L}_{k,i}$ remain dimensionless. For each spin $i$, we employ three operator types:
\begin{align}
\hat{L}_{k=1, i} = \sigma_{+}^{i}, \quad \hat{L}_{k=2,i} = \sigma_{-}^{i}, \quad \hat{L}_{k=3,i} = \sigma_{\alpha}^{i}  \text{ with }\alpha \in {x, z, y}.
\end{align}
Here, $\sigma_{\pm} = (\sigma_x \pm i\sigma_y)/2$ are the standard raising and lowering operators~\citep{ghosh2021fast}. The three noise channels are absorption (during charging), dissipation (during discharging), and local decoherence, characterized by the rate constants $\Gamma_+$, $\Gamma_-$, and $\Gamma_\alpha$, respectively. $\Gamma_+$ is the rate of energy input from the charger, $\Gamma_-$ is the rate of energy loss during discharging, and $\Gamma_\alpha$ is the strength of the local noise. Local noise is modeled using bit-flip ($\sigma_x$), phase-flip ($\sigma_z$), and bit-phase-flip ($\sigma_y$) channels. Each spin in the battery interacts with a bosonic reservoir and local noise aligned along the $x$, $z$, or $y$ direction. All parameters with dimensions of energy, including the Lindblad rates $\Gamma_+$, $\Gamma_-$, and $\Gamma_\alpha$, are normalized by the energy scale $\Delta E$. Since all rates are time-independent, the system evolves under Markovian dynamics as described in the literature~\citep{Huang1987}.

To quantify the energy stored in the quantum battery at a given time $t$, we define the battery energy as $E_B(t) = \mathrm{Tr}\big[ \hat{H}_0 \rho(t) \big]$,
where $\rho(t)$ is the time-evolved density matrix of the spin-chain battery obtained from the master equation dynamics with the initial state chosen as the battery ground state. All energy-related quantities are expressed in units of the normalized energy scale $\Delta E$. The maximum amount of work that can be extracted is characterized by the ergotropy $\mathcal{W}(t)$, defined as
\begin{align}
\mathcal{W}(t) = E_B(t) - \min_{U(t)}  \mathrm{Tr}\big[\hat{H}_0 \rho^p(t)\big],
\end{align}
where the minimization is taken over all unitary operations $U(t)$, corresponding to reversible work-extraction protocols. The passive state $\rho^p(t)$ is obtained by diagonalizing $\rho(t)$ and assigning its eigenvalues $r_n$ in decreasing order to the energy eigenstates $|\epsilon_n\rangle$ of $\hat{H}_0$ arranged in increasing energy, such that $\rho^p(t) = \sum_n r_n |\epsilon_n\rangle \langle \epsilon_n|$, consistent with the standard definition~\citep{allahverdyan2004maximal, farina2019charger, Ukhtary2023}. In addition to stored energy and ergotropy, we also evaluate the average charging power $P_B(t) = E_B(t)/t$, and the extractable power $P_\mathcal{W}(t) = \mathcal{W}(t)/t$, which provide insight into how efficiently energy can be deposited into and extracted from the battery during finite-time operation. Since our quantum battery is modeled as a spin-chain, the battery size is specified by the number of spin sites $N$. Both power quantities are expressed in units of $\Delta E$, while the time variable $t$ in Eq.~\eqref{eq:lindblad} is normalized by $\tau_{\Delta E} = \hbar/\Delta E$. With $\hbar = 1$, this normalization reduces to $t \Delta E$, ensuring a dimensionless timescale. Finally, to characterize the charging process from the perspective of quantum information, we employ three standard metrics:\\ (i) the purity,
\begin{align}
\mathcal{P}(t) = \mathrm{Tr}[\rho^2(t)],\label{eq:pur}
\end{align} (ii) the $\ell_1$-norm of coherence, 
\begin{align}
\mathcal{C}_{\ell_1}(t) = \sum_{i\neq j} |\rho_{ij}(t)|, \label{eq:coh}
\end{align} and (iii) the trace distance, 
\begin{align}
\mathcal{D}(t) = \tfrac{1}{2} \|\rho(t)-\rho(0)\|_1, \label{eq:trd}
\end{align} where $\rho(0)$ is the initial state of the battery. These quantities provide a complementary perspective on the role of coherence, decoherence, and state distinguishability in charging dynamics~\citep{Streltsov2017, Lostaglio2015, nielsen-chuang-2010}.

Let us consider a quantum battery composed of two spin-1/2 particles, described by the Hamiltonian in Eq.~\eqref{eq:HB} with the anisotropy parameter $\gamma \leq 1$, corresponding to the $XYZ$ spin-chain model.  We first consider the case of $N=2$ with the purpose of simplifying the analysis while still capturing the essential features of charging dynamics.  By restricting to the two–spin system, one can obtain analytical insights into stored energy, ergotropy, and the role of anisotropy in the charging process before extending to larger many–body systems. The battery is initially prepared in the ground state of the following Hamiltonian:
\begin{align}
    \hat{H}_0 = \begin{pmatrix}
        \mu_+ & 0 & 0 & \kappa\\
        0 & -\frac{J_z}{4} & J & 0 \\
        0 & J & -\frac{J_z}{4} & 0 \\
        \kappa & 0 & 0 & \mu_-
    \end{pmatrix} ,
    \label{eq:H0}
\end{align}
which can be diagonalized immediately.  The corresponding eigenvalues and eigenvectors are
\begin{align}
\epsilon_{0,3} &= \frac{J_z}{4} \mp \eta, ~~~~|\psi_{0,3}\rangle = N_\mp \Big(\frac{\delta_\mp}{\kappa} |00\rangle + |11\rangle \Big), \notag \\
\epsilon_{1,2} &= -\frac{J_z}{4} \mp J, ~~ |\psi_{1,2}\rangle = \frac{1}{\sqrt{2}} \Big(\pm |01\rangle + |10\rangle\Big), 
\end{align}
where $\mu_\pm = J_z/4 \pm \mathcalboondox{h}, ~\kappa=J \gamma$,~~$\eta = \sqrt{\mathcalboondox{h}^2 + \kappa^2}$,~~$\delta_\mp = \mathcalboondox{h} \mp \eta$ and the normalization constant $N_\mp = [(\delta_\mp^2/\kappa^2 )+1]^{-1/2}$. 

We begin by examining the case where each cycle independently experiences a phase-flip (dephasing) channel. During the charging process, the dynamics of the two-spin system is governed by the master equation in the $z$-direction, as explicitly given:

\begin{align}
&\partial_t \rho_z(t) = -i[\hat{H}(t), \rho_z(t) ] \nonumber \\
& + \Gamma_+ \Big[ (\sigma_+ \otimes I)\rho_z(t)(\sigma_- \otimes I) - \frac{1}{2} \{ (\sigma_- \otimes I)(\sigma_+ \otimes I), \rho_z(t) \}  \nonumber \\
& \quad  \quad +  (I \otimes \sigma_+)\rho_z(t)(I \otimes \sigma_-) - \frac{1}{2} \{ (I \otimes \sigma_-)(I \otimes \sigma_+), \rho_z(t) \}  \Big] \nonumber \\
& + \Gamma_Z \Big[ (\sigma_z \otimes I)\rho_z(t)(\sigma_z \otimes I) + (I \otimes \sigma_z)\rho_z(t)(I \otimes \sigma_z) \\ \nonumber
& \quad \quad - 2\rho_z(t)\Big].
\end{align}
Here, $\hat{H}(t) = \hat{H}_0 + \hat{H}_c$ includes both the intrinsic battery Hamiltonian $\hat{H}_0$ and the charging Hamiltonian $\hat{H}_c$. The initial state is chosen as the ground state of $\hat{H}_0$, so that the system starts in an uncharged configuration. The term proportional to $\Gamma_+$ describes incoherent excitation due to local noise-assisted processes, while the $\Gamma_Z$ term represents pure dephasing in the $z$-direction.

During the discharging process, the external charging field is turned off ($\hat{H}_c \rightarrow 0$), such that the system evolves solely under the intrinsic Hamiltonian $\hat{H}_0$. The master equation then reads:

\begin{align}
&\partial_t \rho_z(t) = -i[\hat{H}_0, \rho_z(t) ] \nonumber \\
& + \Gamma_- \Big[ (\sigma_- \otimes I)\rho_z(t)(\sigma_+ \otimes I) - \frac{1}{2} \{ (\sigma_+ \otimes I)(\sigma_- \otimes I), \rho_z(t) \}  \nonumber \\
& \quad  \quad +  (I \otimes \sigma_-)\rho_z(t)(I \otimes \sigma_+) - \frac{1}{2} \{ (I \otimes \sigma_+)(I \otimes \sigma_-), \rho_z(t) \}  \Big] \nonumber \\
& + \Gamma_Z \Big[ (\sigma_z \otimes I)\rho_z(t)(\sigma_z \otimes I) + (I \otimes \sigma_z)\rho_z(t)(I \otimes \sigma_z) \\ \nonumber
& \quad \quad - 2\rho_z(t)\Big].
\end{align}
In this case, the $\Gamma_-$ term represents local relaxation processes that allow energy stored in the battery to be released to the environment. The dephasing term $\Gamma_Z$ remains active, ensuring that coherence decay continues to influence dynamics even during energy extraction. This formulation allows us to consistently study both charging and discharging dynamics within the same phase-flip noise model, highlighting how local dephasing stabilizes energy storage and affects the extractable work over time.
 
 As the initial state, we choose the eigenstate of the system's Hamiltonian $\hat{H}_0$ with the lowest eigenvalue.
\begin{align}
    \rho_B^{(\uparrow, \downarrow)} = 
    \begin{pmatrix}
        \frac{\delta_\mp^2}{\delta_\mp^2+\kappa^2} & 0 & 0 & \frac{\kappa \delta_\mp}{\delta_\mp^2+\kappa^2} \\
        0 & 0 & 0 & 0 \\
        0 & 0 & 0 & 0 \\
        \frac{\kappa \delta_\mp}{\delta_\mp^2+\kappa^2} & 0 & 0 & \frac{\kappa^2}{\delta_\mp^2+\kappa^2}
    \end{pmatrix}
    \label{eq:dens}
\end{align}
In this setup, both the charger and the noise act locally. The initial state of the battery, $\rho_B^{(\uparrow, \downarrow)}$, is selected as the ground state of the Hamiltonian. This choice ensures that $\rho_B^\uparrow$ supports the charging process, while the excited state $\rho_B^\downarrow$ corresponds to the discharging process.  The simplicity of the $N=2$ model allows us to directly compute the full spectrum of the system and to analyze how the interplay between anisotropy and spin–spin couplings influences the charging efficiency. These results provide a baseline for understanding more complex scenarios with $N > 2$ spins, where collective effects and many–body correlations become more pronounced.

\section{Results and discussion}
In this section, we first investigate the effect of varying the size $N$ of the spin chain to determine the optimal $N$ that balances storage capacity and computational cost.  For this part, we restrict the analysis to a single type of local noise, namely pure dephasing (phase-flip), and the corresponding results are presented in Sec.~\ref{subsec:scaling}.  In the second part, we focus on the time evolution of battery performance to study both charging and discharging processes.  Here, we vary the noise strength to explore the emergence of Zeno-like stabilization effects and consider three different local noise channels: bit-flip, phase-flip, and bit-phase-flip errors. Those results are presented in Sec.~\ref{subsec:time}.
\label{sec:result}

\subsection{Scaling Behavior under Local Noise}
\label{subsec:scaling}

\begin{figure*}[tb]
  \includegraphics[width=0.9\textwidth]{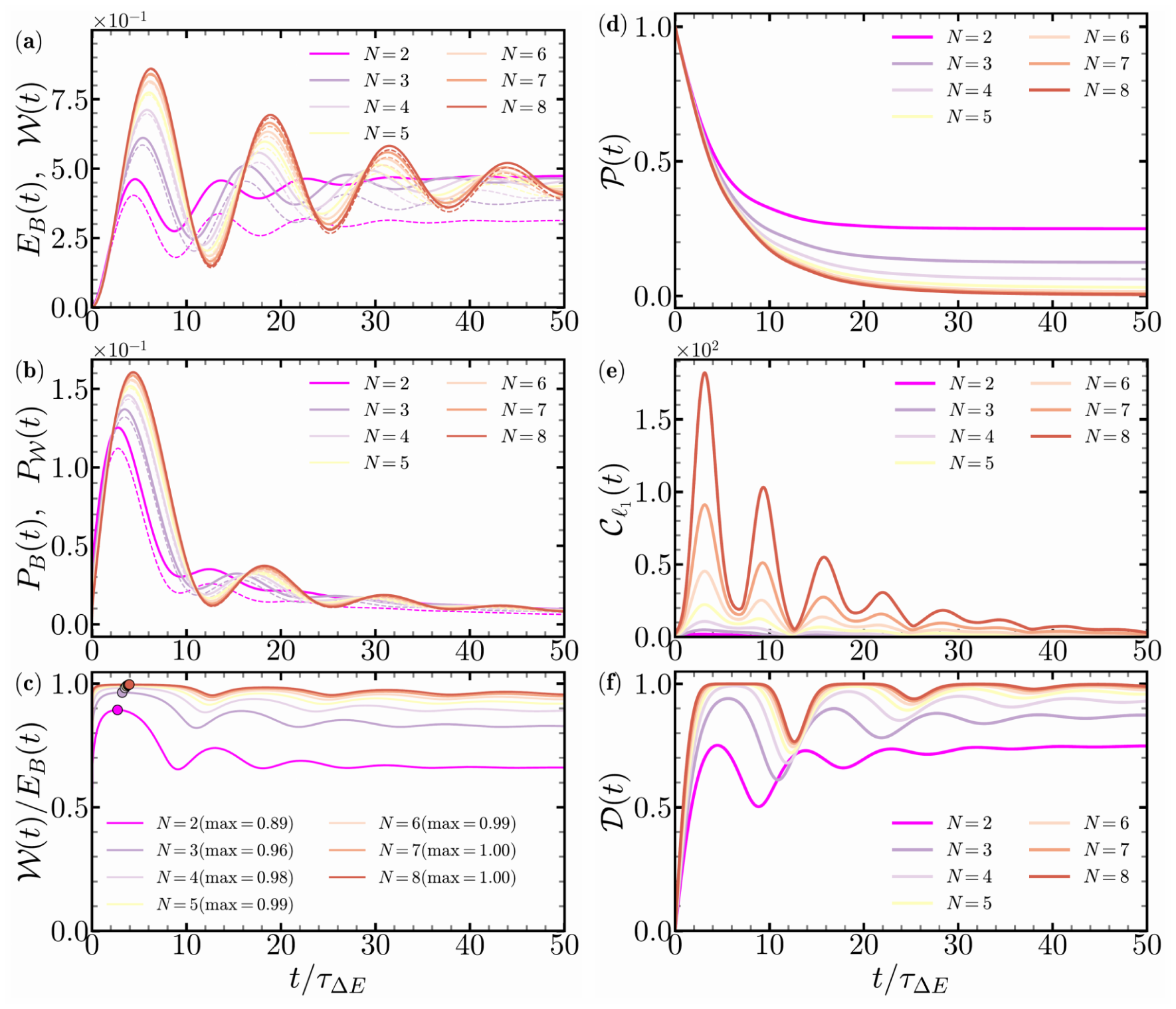}
  {\caption{\justifying Performance of a quantum battery based on an anisotropic $XYZ$ Heisenberg spin chain under a \emph{local charging protocol}. Each spin is driven by a local transverse magnetic field along the $x$-axis ($\omega<1$) and independently coupled to a bosonic bath that induces excitation channels $L_{+,i}=\sqrt{\Gamma_{+}}\,\sigma_{+}^{i}$ (with $\Gamma_{+}/\Delta E=0.01$). We switch off the discharging channel ($\Gamma_{-}/\Delta E=0$) and include local pure dephasing noise $L_{z,i}=\sqrt{\Gamma_{Z}}\,\sigma_{z}^{i}$ (with $\Gamma_{Z}/\Delta E=0.06$). System parameters are fixed at $\lambda
\equiv J/|\mathcalboondox{h}|=0.5$ (paramagnetic regime with dominant field), anisotropy $\gamma=0.5$, and longitudinal coupling $J_{z}=0.2$, all expressed in units of $\Delta E$, while the number of spins $N$ is varied. Panels (a)–(c) depict stored energy $E_B(t)$ and ergotropy $\mathcal{W}(t)$, charging power $P_{B}(t)$ and extractable power, and the ergotropy-to-energy ratio, all as functions of dimensionless timescale $t/\tau_{\Delta E}$.  Panels (d)–(f) show purity $\mathcal{P}(t)$, $l_{1}$-norm coherence $\mathcal{C}_{l_{1}}(t)$, and trace distance $\mathcal{D}(t)$.}}
  \label{fig:QB-N}
\end{figure*}

We begin by analyzing the charging dynamics of the spin-chain quantum battery under local noise, with a particular focus on how the performance scales with the number of constituent spins, $N$. Determining the appropriate system size represents the first crucial step in investigating spin-chain quantum batteries, as it establishes the baseline for understanding the interplay between energy storage capacity, the fraction of extractable energy, and the computational feasibility of simulating many-body dynamics. In this work, we systematically generated spin chains of varying lengths, from $N=2$-8, with Hilbert-space dimension scaling as $2^N$. This approach allows us to probe the effects of system size on both dynamical and steady-state properties in a controlled manner. The exponential growth of the Hilbert space with increasing $N$ immediately underscores a fundamental tradeoff: larger chains can, in principle, store more energy, but they also require rapidly increasing computational resources to evolve the full density matrix, including both memory and numerical time requirements. By performing this systematic survey, we establish a foundation for identifying an optimal chain length that maximizes stored energy and extractability while remaining computationally tractable, which in turn provides guidance for the design of future experimental and theoretical studies.

As shown in Figs.~\ref{fig:QB-N}(a)–(c), local dephasing induces a clear Zeno-like stabilization of the charging dynamics. Figure~\ref{fig:QB-N}(a) shows that the stored energy $E_B(t)$ exhibits an initial rapid growth followed by damped oscillations that converge to a noise-stabilized plateau. The ergotropy $\mathcal{W}(t)$ follows a similar trend but remains systematically below $E_B(t)$, evidencing the presence of a passive contribution, which is significantly more pronounced in small chains ($N=2$–4).  Figure~\ref{fig:QB-N}(b) shows that the charging power $P_B(t)$ and ergotropy rate $P_\mathcal{W}(t)$ display sharp initial peaks, whose amplitudes increase with $N$ but are rapidly suppressed, consistent with the dephasing-induced freezing of coherent revivals.  Most notably, Fig.~\ref{fig:QB-N}(c) demonstrates that the ergotropy-to-energy ratio $\mathcal{W}(t)/E_B(t)$ reaches its maximum at early times and grows with $N$: $N=2$ reaches $\max=0.89$ at $t/\tau_{\Delta E}\approx 2.70$, $N=3$ gives $\max=0.96$ at $t/\tau_{\Delta E}\approx 3.20$, $N=4$ yields $\max=0.98$ at $t/\tau_{\Delta E}\approx 3.50$, $N=5$ achieves $\max=0.99$ at $t/\tau_{\Delta E}\approx 3.65$, $N=6$ reaches $\max=0.99$ at $t/\tau_{\Delta E}\approx 3.80$, while the largest chains, $N=7$ and $N=8$, attain nearly perfect ratios $\max=1.00$ at $t/\tau_{\Delta E}\approx 3.85$ and $3.95$, respectively. For $N=6$, the ratio remains close to unity even at long times, indicating that nearly all deposited energy is extractable as work. In contrast, smaller chains charge faster but store less energy and retain a sizable passive component, whereas $N=8$ maximizes storage capacity at the cost of computational efficiency and practical extractability.

The quantum information metrics displayed in Figs.~\ref{fig:QB-N}(d)–(f) further corroborate this picture. The purity decreases monotonically with $N$, reflecting stronger effective mixing induced by local noise. The $\ell_1$-norm coherence shows an initial transient peak coinciding with the early maxima of charging power, followed by damped oscillations that signify the progressive loss of coherence. Since coherence, defined in Eq.~\eqref{eq:coh}, quantifies the off-diagonal contributions in the energy eigenbasis, it essentially determines the capacity of quantum storage, as large coherence amplitudes correspond to enhanced charging beyond classical limits~\cite{Lostaglio2015, korzekwa2016extraction}. In contrast, the trace distance in Eq.~\eqref{eq:trd} and the purity in Eq.~\eqref{eq:pur} capture, respectively, the departure from the initial state and the degree of state mixing. Together, these quantities dictate how much of the stored energy can be converted into useful work, i.e., the ergotropy, which measures the difference between the energy of the actual state and that of its passive counterpart~\cite{Francica2020}. Importantly, coherence provides an upper bound on the excess energy above the passive limit, thereby linking the amount of coherence to the maximum extractable work.

From a resource-theoretic perspective, these three metrics act as complementary indicators of battery performance: the $\ell_1$-norm coherence diagnoses the storage capacity, the purity reflects the quality of the resource by constraining unitary work extraction, and the trace distance quantifies the stability and irreversibility of the charging process. Collectively, they provide a compact fingerprint of the operational resources of the quantum battery. In this sense, local dephasing enforces a Zeno-like suppression of rephasing dynamics: while coherence decay caps the storage amplitude, the concurrent stabilization revealed by purity and trace distance ensures that the stored energy remains highly extractable.

Taken together, the exponential scaling of the Hilbert space with $N$ illustrates why brute-force density-matrix simulations are restricted to relatively small chains. Within this limitation, $N=6$ emerges as an optimal operating point, offering a favorable balance between capacity, extractable work, and computational feasibility. These findings are consistent with prior studies on noise-assisted transport in quantum networks \cite{Plenio2008}, on Zeno dynamics in open quantum systems \cite{Facchi2008}, and on collective scaling properties of quantum batteries \cite{campaioli_quantum_2017,andolina2018charger}, demonstrating that local dephasing can, counterintuitively, stabilize energy storage and improve the fraction of energy extractable as work in many-body quantum systems.

\subsection{Time Evolution of  Performance Metrics}
\label{subsec:time}

Here, we analyze the dynamical evolution of key performance metrics of the quantum battery under the combined action of unitary driving and local decoherence channels. Such figures of merit include the stored energy, the ergotropy, and the charging power, all evaluated as functions of time $t$, thereby quantifying both the capacity and the efficiency of the charging process. The system evolves under the time-dependent Hamiltonian $\hat{H}(t)=\hat{H}_0+\hat{H}_c$, and the master equation is solved numerically to obtain the full density matrix $\rho(t)$. We systematically vary the noise strengths $\Gamma_\alpha$ with $\alpha = x, z, y$, corresponding to bit-flip, phase-flip, and bit-phase-flip channels, in order to assess their impact on coherent energy transfer. By tuning these rates, we emulate conditions that give rise to a Zeno-like stabilization of the charging dynamics, where stronger dephasing suppresses oscillatory revivals and drives the system toward a noise-stabilized steady state.

In addition to energetic performance, we also evaluate quantum-metric quantities, namely the purity, the $\ell_1$-norm coherence, and the trace distance, during the charging stage governed by $\hat{H}(t) = \hat{H}_0 + \hat{H}_c$. These measures provide complementary insights into how coherence and state distinguishability evolve in the presence of noise. Since the discharging process corresponds to setting $\hat{H}_c \to 0$, the metrics become irrelevant in that regime, as they explicitly rely on the interplay between the free Hamiltonian and the coherent driving term.

\subsubsection{Charging process}
\begin{figure*}[tb]
  \includegraphics[width=0.9\textwidth]{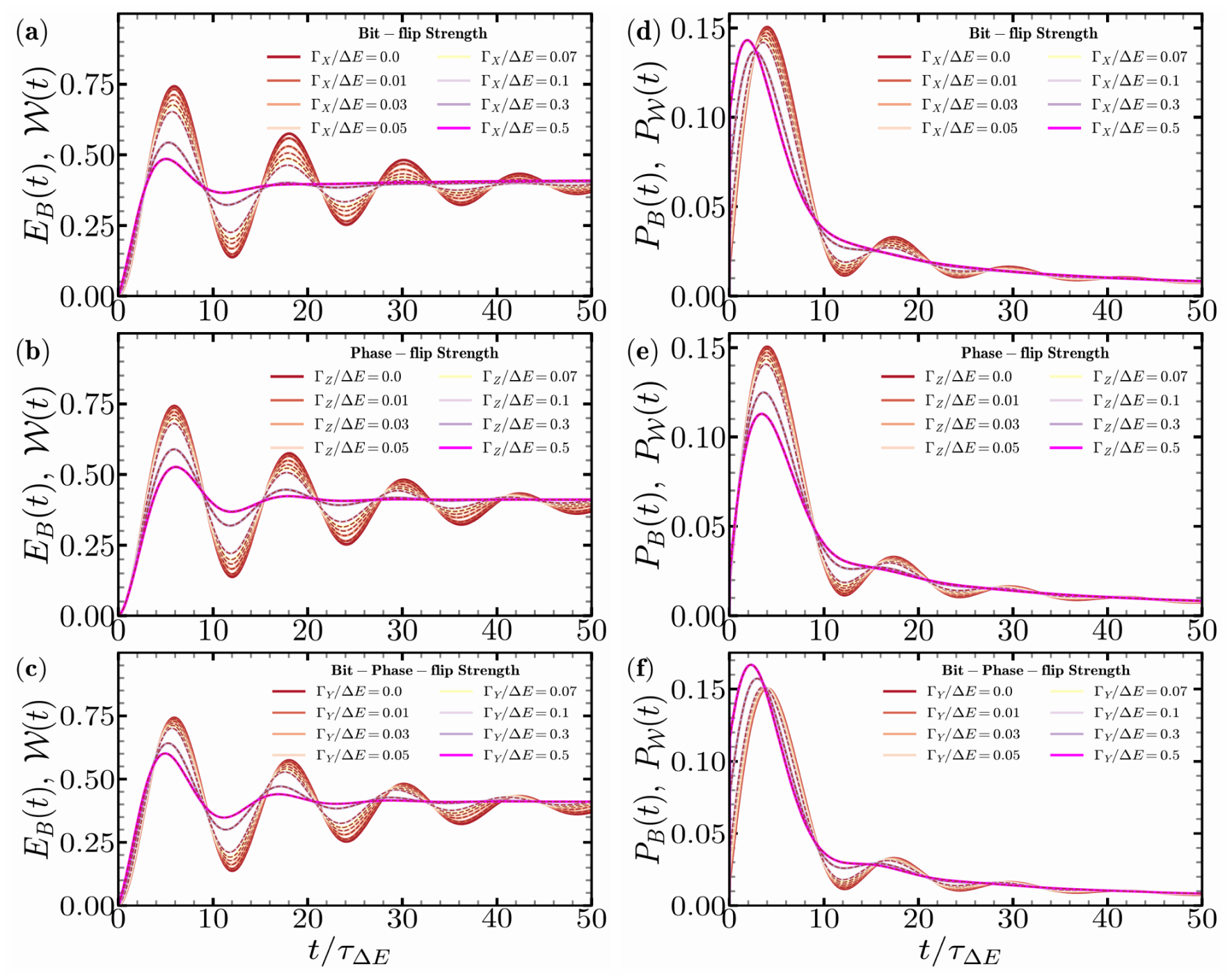}
  {\caption{\justifying
Performance of a quantum battery under local noise channels for $N=6$ spins during the charging process. Panels (a)–(c) show total stored energy $E_{B}(t)$ (solid line) and ergotropy $\mathcal{W}(t)$ (brown dashed line) under bit-flip, phase-flip, and bit-phase-flip noise, respectively, as functions of dimensionless timescale $t/\tau_{\Delta E}$. Panels (d)–(f) show charging power $P_{B}(t)$ (solid line) and extractable power $P_{\mathcal{W}}(t)$ (brown dashed line) under the same noise conditions.  The calculation parameters for this system are as in Fig.~\ref{fig:QB-N}, with $\lambda=0.5$, $\gamma=0.5$, and $J_{z}=0.2$.}}
\label{fig:charging}
\end{figure*}

\begin{figure*}[tb]
  \includegraphics[width=\textwidth]{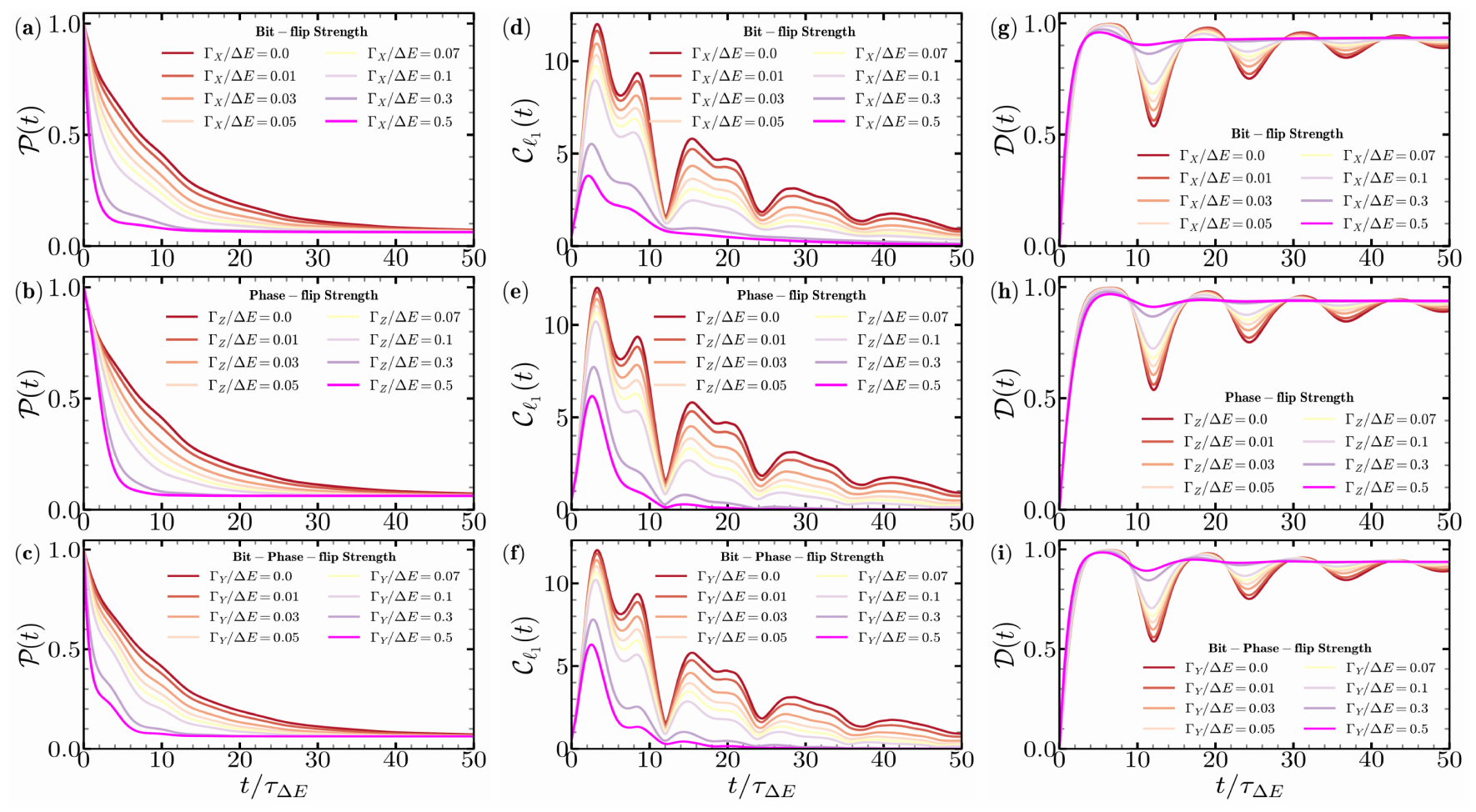}
  {\caption{\justifying
Quantum-information metrics of the quantum battery during the charging process for $N=6$ spins under local noise channels.  (a)–(c)~Purity $\mathcal{P}(t)$. (d)–(f)~$l_{1}$–norm coherence $\mathcal{C}_{l_{1}}(t)$. (g)–(i)~Trace distance $\mathcal{D}(t)$. All quantities are plotted as a function of dimensionless timescale $t/\tau_{\Delta E}$ for three different noise types (bit-flip, phase-flip, and bit-phase-flip).  The system parameters are the same as in Fig.~\ref{fig:QB-N}.}}
  \label{fig:QB-3panel}
\end{figure*}

We initialize the quantum battery in the fully polarized state $\rho_B^{\downarrow}$ [Eq.~\eqref{eq:dens}] and charge it via a coherent drive $\hat{H}c$ with frequency $\omega < 1$, assisted by a local bosonic bath that induces excitation channels
$L_{+,i} = \sqrt{\Gamma_+}, \sigma_+^{i}, \quad \Gamma_+/\Delta E = 0.01,$
while the discharging channels are turned off, $L_{-,i} = \sqrt{\Gamma_-}, \sigma_-^{i}$ with $\Gamma_-/\Delta E = 0$, and all rates are scaled by the energy spacing $\Delta E$. On top of this baseline protocol, we simultaneously include the three canonical local Pauli noise channels: the bit-flip ($\sigma_x$), phase-flip ($\sigma_z$), and bit-phase-flip ($\sigma_y$), modeled via the operator-sum representation~\cite{nielsen-chuang-2010}. Each channel is assigned an independent strength
$\Gamma_\alpha/\Delta E \in \{0, 0.01, 0.03, 0.05, 0.07, 0.1, 0.3, 0.5\},~\alpha = x, z, y,$
allowing us to systematically explore the interplay between coherent charging, dissipative excitation, and multi-channel noise. All energies are scaled by $\Delta E$ and all times by the characteristic timescale $\tau_{\Delta_E} = \hbar/\Delta E$ (with $\hbar=1$).

Figures~\ref{fig:charging}(a)–(c) show the stored energy $E_B(t)$ (solid lines) and the ergotropy $\mathcal{W}(t)$ (brown dashed lines). For bit-flip and phase-flip, the behavior follows the conventional expectation that noise degrades performance: increasing noise lowers the energy plateaus and reduces $\mathcal{W}(t)$, as it washes out the coherent Rabi-like oscillations that feed the charging bursts and mixes populations into more passive configurations. In contrast, Fig.~\ref{fig:charging}(c) shows that under bit–phase-flip a distinct regime emerges: at sufficiently large noise strengths, the curves nearly recover the low-noise performance and $\mathcal{W}(t)$ closely tracks $E_B(t)$. This outcome can be understood because the dephasing component suppresses destructive recurrences, thereby stabilizing the charged state in a Zeno-like manner, while the bit-flip component facilitates population transfer that constructively cooperates with the bath pumping. Such an interplay reflects the well-known mechanism of environment-assisted quantum transport and anti-Zeno acceleration, where noise becomes beneficial when its rate is comparable to the intrinsic system timescales~\cite{rebentrost2009environment,Plenio2008,Maier2019}.

The power curves in Figs.~\ref{fig:charging}(d)–(f) confirm the above notion.  Figure~\ref{fig:charging}(e) (phase-flip) shows that the first peak is shifted to earlier times and its amplitude is strongly reduced, consistent with the damping of coherence that drives constructive interference for fast charging. Figure~\ref{fig:charging}(d) (bit-flip) shows that stronger noise accelerates the arrival of the maximum but the peak value is smaller than in the low-noise case, since rapid mixing prevents the coherent buildup of large currents. Figure~\ref{fig:charging}(f) (bit-phase-flip) shows that the combination of dephasing and population transfer produces an advantageous regime where the power peaks appear earlier and reach large amplitudes, both for the total and extractable power. The near overlap of $\mathcal{W}(t)$ and $E_B(t)$ in this favorable bit–phase-flip regime indicates that the stored energy is dominantly ergotropic, consistent with the ergotropy framework where work extraction is determined by the state’s spectrum relative to the system Hamiltonian.

Having established the charging performance of the quantum battery in Fig.~\ref{fig:charging}, we now examine the time evolution of the quantum information metrics, which provide complementary insight into the role of coherence and noise during the charging process. Figure~\ref{fig:QB-3panel} reports the purity $\mathcal{P}(t)$, the $\ell_{1}$-norm coherence $\mathcal{C}_{\ell_1}(t)$, and the trace distance $\mathcal{D}(t)$ for a battery of $N=6$ spins under local bit–flip, phase–flip, and bit–phase–flip noise channels.  The purity, shown in Figs.~\ref{fig:QB-3panel}(a)–(c), exhibits a monotonic decay whose rate increases with the noise strength, reflecting the progressive mixing of the battery state due to the environment. This degradation correlates with the suppression of coherent contributions to the stored energy and consequently to the ergotropy. Figures~\ref{fig:QB-3panel}(d)–(f) demonstrate that the $\ell_{1}$-norm coherence undergoes damped oscillations, whose amplitude diminishes with increasing noise. Notably, phase-flip noise is particularly detrimental to coherence, as expected from its direct action on off-diagonal terms, while bit–flip noise alters both populations and coherence. Finally, the trace distance in Figs.~\ref{fig:QB-3panel}(g)–(i) captures the transient oscillations driven by the interplay of coherent charging dynamics and dissipation, before relaxing toward steady values determined by the noise channel and its strength.

The behaviors mentioned above highlight the strong connection between coherence, purity, and the battery’s ability to store useful work. The observed decay of $\mathcal{C}_{\ell_1}(t)$ and $\mathcal{P}(t)$ parallels the reduction of ergotropy in Fig.~\ref{fig:charging}, confirming that quantum coherence constitutes a key resource for extractable work in finite-time charging processes~\cite{Francica2020, Kamin2020}. Beyond this, the three metrics provide complementary roles: purity and trace distance are directly linked to the maximal extractability of energy, as they measure how far the state remains from an idealized pure or target configuration; in contrast, the $\ell_{1}$-norm coherence predominantly determines the amplitude of charging oscillations and thus bounds the effective capacity achievable at finite times. The distinct impact of different Pauli channels emphasizes that the type of local noise, not only its strength, critically shapes the charging dynamics.  Together, these results demonstrate that quantum information metrics serve as valuable diagnostics of battery performance, bridging microscopic decoherence dynamics with macroscopic figures of merit such as stored energy, ergotropy, and power. More importantly, they highlight that noise engineering, by selectively suppressing or redirecting decoherence pathways, offers a viable route to tailor the trade-off between charging capacity, efficiency, and extractability. This perspective underscores the central role of coherence as a non-classical resource and establishes purity, coherence, and trace distance as fundamental indicators for optimizing quantum battery protocols under realistic noisy conditions~\cite{ferraro-collectivecharging-2018, ghosh2021fast}.

\subsubsection{Discharging process}

\begin{figure*}[tb]
  \includegraphics[width=0.9\textwidth]{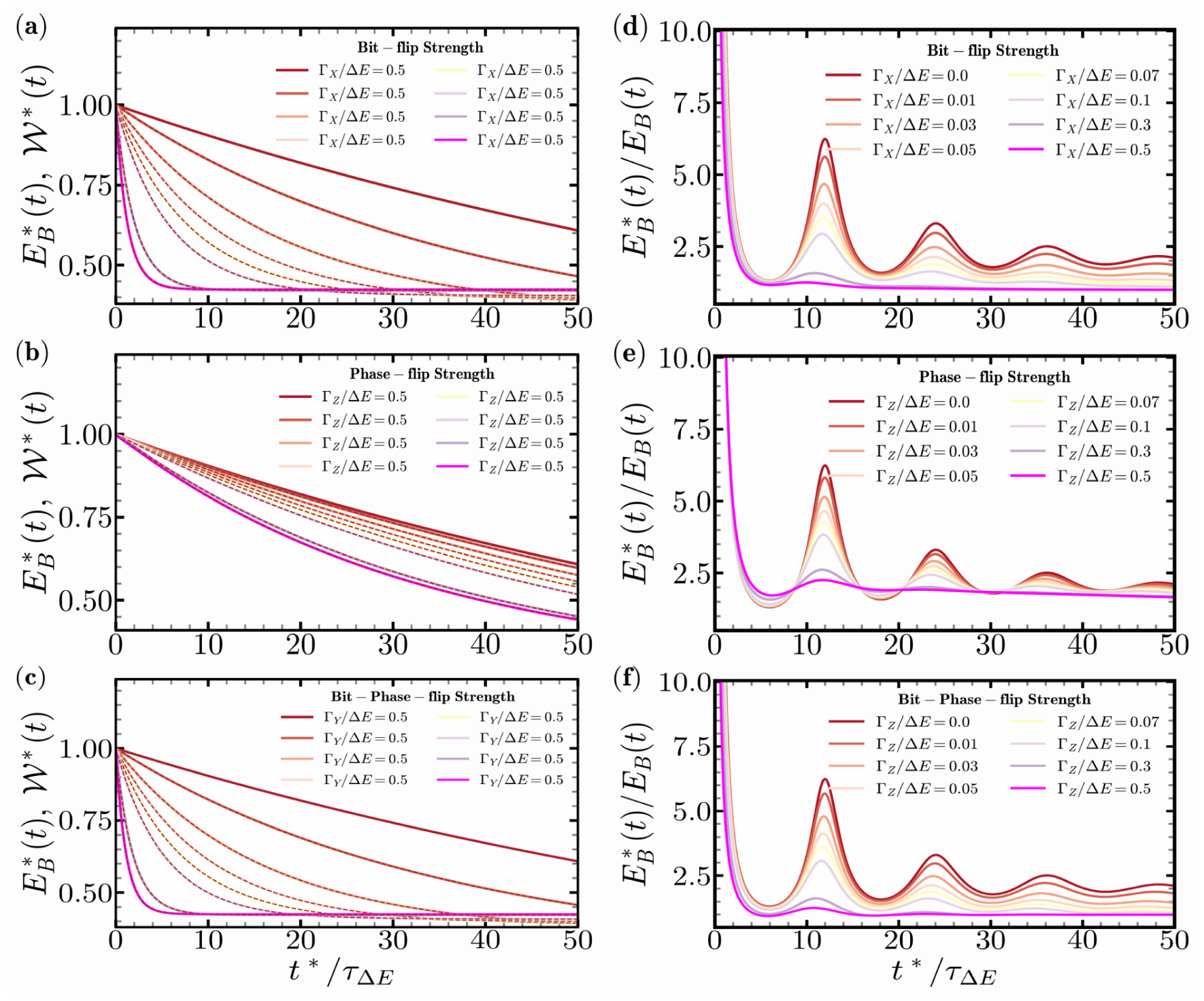}
  {\caption{\justifying
Performance of a quantum battery under local noise channels for $N=6$ spins during the discharging process. Panels (a)–(c) show total released energy $E^*_{B}(t)$ (solid line) and released ergotropy $\mathcal{W}^*(t)$ (brown dashed line) under bit–flip, phase–flip, and bit–phase–flip noise, respectively, as functions of dimensionless timescale $t^*/\tau_{\Delta E}$.  Panels (d)–(f) show the released energy-to-stored energy ratio under the same noise conditions.  System parameters are as in Fig.~\ref{fig:QB-N}, with $\lambda=0.5$, $\gamma=0.5$, and $J_{z}=0.2$.}}
  \label{fig:QB-Discharging}
\end{figure*}

After the charging process, we investigate the release of useful work by initializing the battery in the charged polarized state $\rho_B^{\uparrow}$ (Eq.~\ref{eq:dens}). During the discharging process, the coherent driving term $\hat{H}_c$ (Eq.~\ref{eq:c}) is switched off, so that the total Hamiltonian reduces to the intrinsic battery's Hamiltonian. Energy release occurs through the local bosonic bath, which induces relaxation channels $
L_{-,i} = \sqrt{\Gamma_-}\, \sigma_-^{i}, \quad \Gamma_- / \Delta E = 0.01,
$ while excitation channels are suppressed ($\Gamma_+ = 0$). Analogous to the charging protocol, we include three local Pauli noise channels, bit-flip ($\sigma_x$), phase-flip ($\sigma_z$), and bit-phase-flip ($\sigma_y$), with independent strengths $
\Gamma_\alpha / \Delta E \in \{0,0.01,0.03,0.05,0.07,0.1,0.3,0.5\}, \quad \alpha = x, z, y,
$ to assess their impact on the discharging dynamics~\cite{nielsen-chuang-2010}. The difference from the charging process is that during discharging $L_{-,i}$ is active while $L_{+,i}$ is turned off, whereas during charging $L_{+,i}$ is active and $L_{-,i}$ is suppressed. The time evolution is analyzed in terms of the discharging time $t^*$, which is obtained from the density operator during the relaxation process, $\rho_B^{\downarrow}(t^*)$. For convenience, we normalize the timescale as $ t^*/\tau_{\Delta_E}, \text{ with } \tau_{\Delta_E} = \hbar/\Delta E,
$ so that all dynamics are expressed in dimensionless units. All energies are similarly expressed in units of $\Delta E$.

Figures~\ref{fig:QB-Discharging}(a)–(c) show the time evolution of the released energy $E_B^*(t)$ (solid lines) and the released ergotropy $\mathcal{W}^*(t)$ (brown dashed lines) under the three Pauli channels. In all cases, the total released energy exhibits a monotonic decay, with faster losses for increasing noise strengths. The released ergotropy is consistently lower than the total energy, reflecting that only a fraction of the stored energy can be extracted as useful work, in agreement with general resource-theoretic arguments~\cite{allahverdyan2004maximal,Francica2020}. The gap between $E_B^*(t)$ and $\mathcal{W}^*(t)$ grows with stronger noise, confirming that decoherence reduces the work-extractable component more severely than the raw energy release.  Figures~\ref{fig:QB-Discharging}(d)–(f) display the ratio $E_B^*(t)/E_B(t)$, quantifying the efficiency of the conversion of energy to work during discharge.  The oscillatory patterns at short times reveal a coherent back-action from the system-bath interaction, gradually washed out under stronger noise.  Notably, the phase-flip noise most strongly suppresses efficiency, consistent with its destructive effect on coherence resources identified in the charging stage (Fig.~\ref{fig:charging}). By contrast, bit-flip noise allows for a relatively higher efficiency at intermediate times, though eventually all channels drive the ratio toward lower steady values. These results demonstrate that decoherence not only limits the stored energy, but also strongly suppresses its useful release, highlighting a fundamental asymmetry between charging and discharging in noisy quantum batteries~\cite{ferraro-collectivecharging-2018,Francica2020,ghosh2021fast}.

Beyond the results presented here, several directions remain open. First, while our study has characterized general trends with coupling strength $J$, it remains an open problem to determine whether there exists an optimal interaction strength that minimizes the performance gap between noisy and noiseless cases. Identifying such a regime would clarify how internal couplings and external noise can be co-engineered for robust charging. Second, although we observed that phase-flip noise can stabilize charging dynamics through Zeno-like suppression of rephasing, the possibility of exploiting bit-phase-flip noise for constructive effects requires deeper exploration, especially in relation to many-body correlations and nonlinear dynamics. Finally, extending the analysis to larger system sizes or different network topologies remains an important challenge given the exponential growth of the Hilbert space. Addressing these problems would provide valuable guidance for designing noise-resilient quantum thermodynamic devices~\citep{le-spinchain-2018,ferraro-collectivecharging-2018}.

\section{Conclusions}
\label{sec:conclusion}
We have systematically investigated the charging and discharging dynamics of a spin-chain quantum battery subject to three local noise channels: bit-flip, phase-flip and bit-phase-flip, with the interspin coupling $J$ fixed as part of the model. In the absence of noise, the battery exhibits coherent energy oscillations, with ergotropy closely tracking the total stored energy. The presence of decoherence generally damps these oscillations, although the specific effect depends on the type of noise. Bit-flip noise proves to be the most detrimental, accelerating charging while strongly limiting both capacity and ergotropy. Phase-flip noise slows charging, but allows stored energy and ergotropy to persist longer during discharging. Remarkably, bit-phase-flip noise exhibits a Zeno-like behavior: In the high-noise-strength regime, it enhances charging speed and stabilizes stored energy and ergotropy, whereas at low noise strengths, it degrades performance.

These results highlight that optimal operation of quantum batteries in open systems requires the balance of fast charging, energy stability, and controlled decoherence. An important open problem concerns the role of interspin coupling $J$: Although its fixed value influences energy transport and correlations, it remains unclear whether certain coupling regimes could mitigate noise-induced degradation while enhancing charging performance.  Exploring this interaction offers a promising avenue for designing high-performance and noise-resilient quantum batteries.
\begin{acknowledgments}
The authors thank Mahameru BRIN for allowing us to utilize their HPC facilities.  We also acknowledge Dr. M~Shoufie Ukhtary, Dr. Emir S. Fadhilla, and Dr. Eddwi H. Hasdeo (all at the BRIN Research Center for Quantum Physics) who always stimulate fruitful discussion in our regular group meetings.  One of the authors, M.M.F., is supported by a research assistantship from the BRIN Directorate for Talent Management.  The codes and data supporting the findings of this study are available from the corresponding authors upon reasonable request.  
\end{acknowledgments}

%
\end{document}